\documentclass[11pt]{article}
\usepackage{amssymb}
\usepackage{amsmath}
\usepackage{amscd}
\usepackage{latexsym}
\usepackage{bbm}

\catcode`@=11
\renewcommand{\section}{\@startsection{section}{1}{0pt}{\medskipamount}
{\medskipamount}{\large\bf}}
\numberwithin{equation}{section}
\catcode`@=12
\def\a{\alpha}
\def\b{\beta}

\def\de{\delta}
\def\eps{\epsilon}
\def\ve{\varepsilon}

\def\vp{\varphi}

\def\O{\Omega}

\newcommand{\C}{\mathbb C}
\newcommand{\R}{\mathbb R}

\newcommand{\Gcal}{{\cal G}}
\newcommand{\Acal}{{\cal A}}

\newcommand{\Mcal}{{\cal M}}
\newcommand{\Fcal}{{\cal F}}
\newcommand{\Ncal}{{\cal N}}

\newcommand{\Z}{{\mathbb Z}}
\newcommand{\gfrak}{{\mathfrak g}}

\def\im{\textrm{i}}
\def\tr{\textrm{tr}}
\def\diff{\textrm{d}}
\def\pa{\mbox{$\partial$}}
\def\sfrac#1#2{{\textstyle\frac{#1}{#2}}}
\def\+{\dagger}
\def\={\ =\ }
\def\und{\qquad\textrm{and}\qquad}
\def\and{\quad\textrm{and}\quad}
\def\with{\quad\textrm{with}\quad}
\def\for{\quad\textrm{for}\quad}

\def\Id{\mathrm{Id}}

\begin{document}

\begin{titlepage}
\setcounter{page}{0}

\hspace{2.0cm}

\begin{center}

{\LARGE\bf
Non-Abelian sigma models from Yang--Mills theory \\[5mm] 
compactified on a circle}

\vspace{12mm}

{\Large
Tatiana A. Ivanova${}^*$, \ Olaf Lechtenfeld${}^{\+\times}$ \ and \  Alexander D. Popov${}^\+$
}\\[10mm]
\noindent ${}^*${\em
Bogoliubov Laboratory of Theoretical Physics, JINR\\
141980 Dubna, Moscow Region, Russia
}\\
{Email: ita@theor.jinr.ru
}\\[5mm]
\noindent ${}^\+${\em
Institut f\"ur Theoretische Physik, 
Leibniz Universit\"at Hannover \\
Appelstra\ss{}e 2, 30167 Hannover, Germany
}\\
{Email: alexander.popov@itp.uni-hannover.de}
\\[5mm]
\noindent ${}^\times${\em
Riemann Center for Geometry and Physics,
Leibniz Universit\"at Hannover \\
Appelstra\ss{}e 2, 30167 Hannover, Germany
}\\
{Email: olaf.lechtenfeld@itp.uni-hannover.de}

\vspace{20mm}

\begin{abstract}
\noindent 
We consider SU($N$) Yang--Mills theory on $\R^{2,1}\times S^1$, where $S^1$ is a spatial circle. 
In the infrared limit of a small-circle radius the Yang--Mills action reduces to the action 
of a sigma model on $\R^{2,1}$ whose target space is a $2(N{-}1)$-dimensional torus modulo the 
Weyl-group action. We argue that there is freedom in the choice of the framing of the gauge bundles, 
which leads to more general options. In particular, we show that this low-energy limit can give rise to 
a target space SU$(N){\times}$SU$(N){/}\Z_N$. The latter is the direct product of SU($N$) and its 
Langlands dual SU$(N){/}\Z_N$, and it contains the above-mentioned torus as its maximal Abelian 
subgroup. An analogous result is obtained for any non-Abelian gauge group.

\end{abstract}

\end{center}
\end{titlepage}

\section {Introduction and summary} 

\noindent
Pure Yang--Mills or QCD-like theories in four spacetime dimensions are strongly coupled in the infrared limit. 
It is known that one can partially overcome this difficulty by compactifying Yang--Mills theory on a circle $S^1_R$ with small radius $R$ 
(see e.g.~\cite{1, 2} and references therein). In the adiabatic limit, when the metric on $S^1_R$ is scaled down, 
the $d{=}4$ SU($N$) Yang--Mills action can be reduced (already on the classical level) to the action of a $d{=}3$
sigma model whose target space is $T{\times}T^{\vee}/W$. Here, $T=\;$U(1)$^{N-1}$
is the Cartan torus in  SU($N$) corresponding to Wilson loops around  $S^1_R$, and $T^{\vee}$ is the Cartan torus 
in the Langlands dual SU$(N){/}\Z_N$. The torus $T^{\vee}$ parametrizes the dual (magnetic) photons on $\R^{2.1}$ and corresponds to 't~Hooft
loops around $S^1_R$~\cite{3, 4}. Finally, $W$ is the Weyl group, which for SU($N$) is the finite permutation group $S_N$.

The above-mentioned action on $\R^{2,1}$ may be augmented by an effective potential for the sigma-model scalar fields, 
which appears from an additional center-stabilizing term breaking SU($N$) to U(1)$^{N-1}$ and from quantum loop corrections, 
as discussed e.g.\ in~\cite{5a,5}. In our paper we focus on the derivation of kinetic terms in the low-energy limit of pure Yang--Mills theory.
Therefore, for the time being, we ignore a possible symmetry-breaking potential.

The main message of the paper is that there is important freedom in the choice of the framing~\footnote{
A bundle over a manifold $M$ is called framed over a submanifold $N\subset M$ if its fibres over $N$ are fixed.
Framed bundles are often used in discussions of instantons and monopoles as well as  on manifolds with boundaries, marked points, punctures etc.\
(see e.g.~\cite{6, 7, 8}).}
of the gauge bundle, and that this leads to the option of enlarging the sigma-model target space from $T{\times}T^{\vee}/W$ to a non-Abelian group, 
up to the maximal space $\Mcal=\;$SU$(N){\times}$SU$(N){/}\Z_N$.
In other words, we shall show how the classical Yang--Mills model on $\R^{2,1}\times S^1_R$ can be reduced to a sigma model on $\R^{2,1}$ 
with non-Abelian target space $\Mcal$ or a subgroup thereof including the torus $T{\times}T^{\vee}\cong\;$U(1)$^{2(N-1)}$. 
For a general gauge group $G$ with weight lattice $\Gamma_w$, the sigma-model target space will be $\Mcal =G{\times}G^{\vee}$, 
where $G^{\vee}$ denotes the Langlands dual group, whose weight lattice $\Gamma^{\vee}_w$ is dual to $\Gamma_w$. Thus, the 
target-space geometry $\Mcal$ of our sigma models obtained from Yang--Mills theory on $\R^{2,1}\times S^1_R$ in the small-$R$ limit 
essentially depends on conditions imposed on the gauge potential~$\Acal$ and the gauge transformations along~$S^1_R$.
 
The Yang--Mills reduction to the Abelian sigma model on $\R^{2,1}$ (where $\cal M$ is toroidal) points at an Abelian confinement mechanism 
based on Dirac monopoles, Abelian vortices and the dual Meissner effect. The Abelian dual superconductor approach has various limitations,
like any other confinement mechanism (see e.g.~\cite{1, 9}). For this reason there have been efforts to extend the dual superconductor mechanism 
to models with non-Abelian monopoles and  non-Abelian vortices (see e.g.~\cite{1, 9, 10} and references therein).
The suggestion of this paper aims in the same direction.

%\newpage

\section {Action functional} 

\noindent {\bf Space $\R^{2,1}\times S^1_R$.}
We consider Yang--Mills theory on the direct product manifold $\R^{2,1}\times S^1_R$ with coordinates
$(x^{\mu})=(x^a, x^3)$, where $x^a\in \R^{2,1}$ and $x^3\in[0,2\pi]$, in which the metric reads 
\begin{equation}\label{2.1}
 \diff s^2_R \= g_{\mu\nu}^R\,\diff x^\mu \diff x^\nu \= \eta_{ab}\,\diff x^a \diff x^b + R^2(\diff x^3)^2\ ,
\end{equation}
where $(\eta_{ab})={\rm diag} (-1,1,1)$ with $a,b=0,1,2$, and the angular coordinate obeys $x^3\sim x^3 +2\pi$.
The dimensionful coordinate $\tilde x^3=Rx^3\sim \tilde x^3+ 2\pi R$ parametrizes the circle $S^1_R$ of radius $R$.

As Yang--Mills structure group we consider mainly $G=\;$SU$(N)$, however an arbitrary semisimple compact Lie group $G$
will also be discussed. Let $I_i$ with $i=1,\ldots,N^2{-}1$ be a basis of the Lie algebra $su(N)$ realized as $N{\times}N$ matrices 
(fundamental representation). We use the normalization condition
\begin{equation}\label{2.2}
\tr \, I_iI_j = -\sfrac12\de_{ij}\  .
\end{equation}
For generators $I_i$ in the adjoint representation of $G$ we will use the same normalization (\ref{2.2}) but with $i=1,\ldots,{\rm dim} G$.

\medskip

\noindent {\bf Gauge fields.}
Let us consider the principal SU$(N)$-bundle $P$ over $\R^{2,1}\times S^1_R$ and the associated complex vector bundle $E\to\R^{2,1}\times S^1_R$
with fibres $V=\C^N$. Let $\Acal$ be a gauge potential (a connection on $P$ and $E$) with values in $su(N)$, so that 
\begin{equation}\label{2.3}
\Fcal \= \diff\Acal + \Acal\wedge\Acal\= \sfrac12\Fcal_{\mu\nu}\,\diff x^\mu \wedge \diff x^\nu\quad\with\quad 
\Fcal_{\mu\nu} \=\partial_\mu\Acal_\nu - \partial_\nu\Acal_\mu + [\Acal_\mu , \Acal_\nu]
\end{equation}
is the $su(N)$-valued gauge field (curvature of $\Acal$). On $\R^{2,1}\times S^1_R$ we have the obvious splitting
\begin{equation}\nonumber
\Acal \=\Acal_{\mu}\,\diff x^\mu\= \Acal_{a}\,\diff x^a+\Acal_{3}\,\diff x^3\ ,
\end{equation}
\begin{equation}\label{2.4}
\Fcal \=\sfrac12\Fcal_{\mu\nu}\,\diff x^\mu \wedge \diff x^\nu \=
\sfrac12\Fcal_{ab}\,\diff x^a \wedge \diff x^b + \Fcal_{a3}\,\diff x^a \wedge \diff x^3\ .
\end{equation}

For unit radius $R{=}1$, indices of $\Fcal_{\mu\nu}$ are raised by the non-deformed inverse metric tensor $(g^{\mu\nu})=(\eta^{ab},1)$.
On the deformed space $\R^{2,1}\times S^1_R$, however, we must raise them with the metric (\ref{2.1}), 
and thus the contravariant field components are
\begin{equation}\label{2.5} 
 \Fcal^{ab}_R = g_R^{ac} g_R^{bd} \Fcal_{cd} = \eta^{ac}\eta^{bd}\Fcal_{cd} = \Fcal^{ab} \und 
 \Fcal^{a3}_R = g_R^{ac}g_R^{33}\Fcal_{c3} = \eta^{ac} R^{-2}\Fcal_{c3} = R^{-2}\Fcal^{a3}\ .
\end{equation}

\medskip

\noindent {\bf Action.} 
We consider the standard Yang--Mills action functional
\begin{equation}\label{2.6}
S\=-\frac{1}{2e^2}\int_{\R^{2,1}\times S^1} \!\!\!\!\!\!\diff^4x\ \sqrt{|\det g^R|}\,\tr\Fcal_{\mu\nu}\Fcal^{\mu\nu}_R
\= -\frac{1}{2e^2R}\int_{\R^{2,1}\times S^1} \!\!\!\!\!\!\diff^4x\ \tr\bigl( R^2\,\Fcal_{ab}\Fcal^{ab}+2\Fcal_{a3}\Fcal^{a3}\bigr)\ ,
\end{equation}
where $e$ is the gauge coupling constant. Here we used (\ref{2.5}) as well as $\det (g_{\mu\nu}^R)=-R^2$.
We do not consider the topological $\theta$-term since finally it will only change the metric on the moduli space.

\section{Adiabatic approach}

\noindent {\bf ``Slow'' and ``fast'' variables.} The adiabatic approach to differential equations, based on the introduction 
of ``slow'' and ``fast'' variables, exists for more than 90 years and is used in many areas of physics. Briefly, if ``slow'' variables
parametrize a space $X$ and  ``fast'' variables parametrize a space $Y$ (of dimensions $p$ and $q$, respectively) then on the
direct product manifold $Z=X\times Y$ one should consider a metric
\begin{equation}\label{3.1}
g_\ve = g_X + \ve^2g_Y\ ,
\end{equation}
where $g_X$ is a metric on $X$,  $g_Y$ is a metric on $Y$ and $\ve\in [0, \infty )$ is a real parameter. The adiabatic limit refers 
to the geometric process of shrinking a metric in some directions while leaving it fixed in the others, $g_X$ in the case  (\ref{3.1}). 
That is, one studies differential equations on $Z=X\times Y$ with the metric  (\ref{3.1}) and the small-$\ve$ limit in 
these equations. More generally, the adiabatic method applies to a fibration $Z\to X$ or if $X$ is a calibrated submanifold of $Z$ (see e.g.~\cite{11}).

\medskip

\noindent {\bf Slow soliton dynamics.} In the simplest case $X=\R$ with $g_X=-1$ (time axis) one looks at solutions of differential 
equations on $Y$ (``static'' solutions) and then switches on a ``slow'' dependence on time. By using this approach, Manton has shown~\cite{12} 
that, in the ``slow-motion limit'', monopole dynamics in Minkowski space $\R^{3,1}=\R^{0,1}\times\R^{3,0}=X\times Y$
can be described by geodesics in the moduli space $\Mcal^n_Y$ of static $n$-monopole solutions. In other words, it was shown~\cite{12, 6, 13}  
that the Yang--Mills--Higgs model  on  $\R^{3,1}$ for slow motion reduces to a sigma model in one dimension whose target space is the $n$-monopole 
moduli space $\Mcal^n_Y$ of solutions to the Yang--Mills--Higgs equations on  $Y=\R^3$.

On three-dimensional manifolds $Y$ with a boundary $\partial Y$, instead of monopoles one may consider nontrivial flat connections and the slow 
dynamics of Chern-Simons ``solitons''; this was done in~\cite{14, 15}. The adiabatic approach was also extended to vortices in $1+2$ dimensions,
to Seiberg-Witten equations in $d{=}4$ Euclidean dimensions, and to instantons viewed as moving solitons in $d=q{+}1\ge 5$ dimensions (see e.g.~\cite{13,16,17,18}
for reviews and references).

\medskip

\noindent {\bf Sigma models on the space $X$ of slow variables.} As far as we know, the adiabatic reduction of (super-)Yang--Mills theory in $p{+}q\ge 4$ dimensions
with $q\ge 2$ to sigma models in $p\ge 2$ dimensions has been investigated for the first time in the physics literature in~\cite{19,20,21} and in the mathematical literature in~\cite{8,22,23}.
The case $q{=}1$ with $Y=S^1$ was studied in~\cite{3,4}, where $\Ncal{=}\,2$ and  $\Ncal{=}\,4$ supersymmetric gauge theories on $\R^{2,1}\times S^1_R$ were reduced to sigma 
models on $\R^{2,1}$.
 
\medskip

\noindent {\bf General scheme of adiabatic reduction.}  For Yang--Mills equations on a $(p{+}q)$-dimensional manifold $X\times Y$ with a metric (\ref{3.1}), 
adiabatic reduction implies the following sequential steps:

\noindent
1) One classifies the Yang--Mills solutions on $Y$ not depending on the coordinates on $X$ and describes the moduli space $\Mcal_Y$ of such solutions. For $q{=}1$
one should consider flat connections on $Y$.

\noindent
2) One assumes that the gauge potential $\Acal =\Acal_X + \Acal_Y$ has  $\Acal_X\ne 0$ and  
that $\Acal$ depends on the coordinates $x^a$ of $X$ only via the moduli-space $\Mcal_Y$ coordinates $X^i$, 
i.e.~by allowing for $X^i=X^i(x^a)$ in $\Acal(X^i)$.

\noindent
3) One substitutes $\Acal =\Acal_X + \Acal_Y$ into the Yang--Mills action functional on $X\times Y$ with the metric (\ref{3.1}) and performs its small-$\ve$ limit.
Then one shows that Yang--Mills theory on $X\times Y$ reduces to a sigma model describing maps from $X$ into the moduli space $\Mcal_Y$. 
For $p{=}3$ and $q{=}1$, the target space is enhanced to $\Mcal_X\times \Mcal_Y$,
where $\Mcal_X$ denotes the moduli space of dual gauge fields on $X$, to be discussed later.

We emphasize that the geometry of the moduli space $\Mcal_Y$ depends essentially on the details of the bundles and connections involved. 
For instance, for flat connections on two-dimensional
manifolds, the geometry of the moduli space $\Mcal_Y$  depends on a boundary (if any) of $Y$, on the number of marked points and punctures, on what kind of bundles is considered (irreducible
or reducible, framed or unframed, with holomorphic or parabolic structure, etc.) and so on  (see e.g.~\cite{7, 8, 24, 25, 26}). As far as we know, for $Y=S^1$ only reducible bundles
with tori as moduli spaces were considered  (see e.g.~\cite{3,4,2,5}). However, the case of flat connections on  two-dimensional spaces $Y$ mentioned above shows that one can have more than
one possibility, that more general cases may be considered. This is what we want to discuss below.

\section{Connections on $S^1$ and their holonomy}

\medskip

\noindent {\bf $G$-bundles over $S^1$.} Let $G$ be a semisimple compact Lie group,\footnote{
Here we consider instead of SU($N$) a semisimple compact Lie group $G$ since this does not change the discussion.} 
$\gfrak$ its Lie algebra, $P= S^1\times G\to S^1$ be a trivial principal $G$-bundle over $Y=S^1$ and $\Acal_{S^1}$ a connection one-form on $P$. 
It will be convenient to parametrize the unit circle $S^1$ by $\exp(2\pi\im\vp )\in S^1$ with $\vp= x^3/2\pi\in [0,1]$. 
The connection $\Acal_{S^1}$ belongs to the space $\Ncal_{S^1}:=\Omega^1(S^1,\gfrak )$ of one-forms on $S^1$ with values in $\gfrak$.\footnote{
We identify $\gfrak$ and its standard dual $\gfrak^*$ using the Killing-Cartan form on $\gfrak$ which is proportional to (\ref{2.2}) for $\gfrak=su(N)$.} 
The loop group $LG=C^{\infty} (S^1, G)$ of gauge transformations in $P$ acts on $\Ncal_{S^1}$ by the formula
\begin{equation}\label{4.1}
f\in LG : \Acal^{}_{S^1}\mapsto \Acal^{f}_{S^1}= f^{-1} \Acal^{}_{S^1} f + f^{-1}\breve\diff f \quad\with\quad
\breve\diff=\diff x^3\pa_3=\diff\vp\pa_\vp\ .
\end{equation}
Note that $f(\vp{+}1)=f(\vp)$ for $f(\vp )\in LG$ (periodicity).

 \medskip

\noindent {\bf Holonomy map.} For any $\vp\in [0,1]$ we introduce the map
\begin{equation}\label{4.2}
h_\vp :\quad \Ncal_{S^1}\ni  \Acal^{}_{S^1}\quad \mapsto\quad  h_\vp(\Acal^{}_{S^1})\in G
\end{equation}
which is defined as the unique solution to the differential equation~\cite{26}
\begin{equation}\label{4.3}
h^{-1}_\vp(\Acal^{}_{S^1})\,\breve\diff h_\vp(\Acal^{}_{S^1})= \Acal^{}_{S^1} \quad\with\quad h_0(\Acal^{}_{S^1})=\Id\ .
\end{equation}
For this map we have the invariance condition
\begin{equation}\label{4.4}
h_\vp(\Acal^{f}_{S^1})=f^{-1}(1)h_\vp (\Acal^{}_{S^1}) f(\vp)\ ,
\end{equation}
where $\Acal^{f}_{S^1}$ is given in (\ref{4.1}). The map  (\ref{4.2}) assigns to any $\Acal^{}_{S^1}\in \Ncal^{}_{S^1}$
a section $(\vp , h_\vp)\in S^1\times G$ of the $G$-bundle $P=S^1\times G$ over $S^1$. 
Note that $h_\vp$ {\it is not periodic\/} in $\vp$, i.e.~$h_0\ne h_1$, since  $h_\vp$ 
defines a line segment in the group $G$
which covers $S^1$ in the base of fibration $P\to S^1$. In fact, $h_1$ is the holonomy of $\Acal^{}_{S^1}$ defining a Wilson loop around  $S^1$.

Recall that the {\it based\/} loop group $\Omega G\subset LG= \Omega G\rtimes G$ is defined as the kernel of the evaluation map 
$LG\to G, f(\vp )\mapsto f(1)$, and therefore $\Omega G = LG/G$. At the endpoint $\vp{=}1$ we get the {\it holonomy map}
\begin{equation}\label{4.5}
h_1 : \quad \Ncal^{}_{S^1}\to G\ ,
\end{equation}
where $h_\vp$ is defined by (\ref{4.3}). From (\ref{4.3}) and (\ref{4.4}) one sees that the action of
$\Omega G$ on $ \Ncal^{}_{S^1}$ is free,\footnote{Recall that $f(1)=\Id$ for $f\in \Omega G$.}
\begin{equation}\label{4.6}
h_1(\Acal^{f}_{S^1})=h_1(\Acal^{}_{S^1})\quad\Leftrightarrow\quad \Acal^{f}_{S^1}(1)=\Acal^{}_{S^1}(1)\for f\in\O G\ ,
\end{equation}
and the holonomy map (\ref{4.5}) is injective on the quotient of  $ \Ncal^{}_{S^1}=\O^1(S^1, \gfrak )$ by $\O G$.
Thus, (\ref{4.5}) is the projection in the principal  $\O G$-bundle over $G$, and  $\Ncal^{}_{S^1}$ is the total space of this bundle~\cite{26}. 
 
\medskip

\noindent {\bf Abelianization.} Consider now the holonomy element $h_1(\Acal^{}_{S^1})\in G$ which parametrizes a flat connection $\Acal^{}_{S^1}$.
From (\ref{4.4}) we see that under gauge transformations $f\in LG$ it is transformed as
\begin{equation}\label{4.7}
h_1(\Acal^{f}_{S^1})=f^{-1}(1)h_1(\Acal^{}_{S^1})f(1)\ ,
\end{equation}
i.e.~only global gauge transformations defined by constant matrices $f(1)\in G$ act on  $h_1(\Acal^{}_{S^1})$. It is known that by a suitable choice of 
$f(1)\in G$ one can transform any element $h_1(\Acal^{}_{S^1})\in G$ to an element in $T/W\subset G$, where $T$  is a maximal 
torus (Cartan torus) in $G$ and $W$ is the Weyl group of $G$. The moduli space of $\Acal^{}_{S^1}$ is defined as the quotient of  $\Ncal^{}_{S^1}$
under the action of the group $LG$ of gauge transformations. In this case, the gauge orbits are parametrized by the orbifold $T/W$.
This is usually meant by ``Abelianization''.

Although the reduction of the group $G$ to its maximal Abelian subgroup is quite popular, many papers claim that it is not natural  
and not even obligatory (see e.g.~\cite{1,9,10} and references therein). We join these arguments by suggesting to control Abelianization 
through the framing of bundles.\footnote{
It is possible to make the reduction from $G$ to $T\subset G$ dependent on extra conditions or parameters (see below).} 

\medskip

\noindent {\bf Framed bundles and moduli of $\Acal^{}_{S^1}$.} Recall that a bundle $E$ over a manifold $Y$ is called framed 
over a point $p\in Y$ if its fibre $E_p$ over this point is fixed  and therefore cannot be transformed by gauge transformations. 
This means that matrices $f$ of gauge transformations at this point are restricted to the identity, $f(p)=\Id$, i.e.~the group $\Gcal$
of gauge transformations in the bundle $E$ reduces to the subgroup $\Gcal_0$ which keeps $E_p$ unchanged. Framing a principal $G$-bundle $P$
over $Y$ at a point $p\in Y$ is achieved by simply fixing  a point $h_p$ in the fibre $G_p$ over $p$.
For instantons on $Y=\R^4$, bundles are framed at infinity in $\R^4$, which forbids global gauge transformations and renders the instanton moduli space 
hyper-K\"ahler. Similarly, for monopoles on $\R^3$, bundles are framed at infinity in $\R^3$, which prevents global Abelian gauge 
transformations generated by the Cartan torus $T$ in the gauge group $G$.  Only after this framing one obtains a hyper-K\"ahler structure
on the moduli space of monopoles. In both cases of instantons and monopoles, the use of unframed bundles is not natural since
the hyper-K\"ahler structure on their moduli spaces is important for various calculations and theoretical predictions. 
In the same spirit, we suggest to frame our $G$-bundles over $S^1$ at the point $\vp =1$ on $S^1$. Then (\ref{4.1})-(\ref{4.7}) imply that
after framing one cannot transform via (\ref{4.7}) the holonomy element $h_1(\Acal^{}_{S^1})$ to the Cartan subgroup
$T\subset G$, since global gauge transformations are no longer allowed. The admissible gauge transformations now belong to the {\it based\/} 
loop group $\O G$, and the moduli space  $\Mcal^{}_{S^1}$ of flat connections $\Acal^{}_{S^1}$ on $P\to S^1$ is therefore the entire group manifold~$G$.

\medskip

\noindent {\bf Dependence on $R$.} The radius $R$ of the circle $ S^1_R$ is a {\it free\/} external parameter. Hence, one can in principle introduce
a dependence on $R$ in the coordinates $X^i$  on $\Mcal^{}_{S^1}\cong G$ in such a way that for $R<R_0$ the group $G$ is reduced to 
some closed subgroup $H\subset G$ (contraction) containing~$T$ and for $R\ge R_0$ one has the whole group $G$. Here $R_0$ is some fixed scale parameter. 
By engineering a suitable dependence $X^i=X^i(R)$, scenarios may be envisioned which are more refined than those in the literature for gauge models 
compactified on $S^1_R$.

\section{Sigma-model effective action} 
 
\noindent
In this final section we consider a gauge group $G$ having in mind $G=\;$SU($N$) with generators $I_i$ and trace (\ref{2.2}). However, one can easily 
generalize all formul\ae\ to a compact semisimple Lie group $G$ by introducing a proper trace normalized as (\ref{2.2}). Then in $\Acal =\Acal^i I_i$ and $\Fcal =\Fcal^i I_i$ 
one can take $I_i$ as generators of SU($N$) or as generators of $G$. Thus, we discuss the generic case and keep $G=\;$SU($N$) as an illustration. 
 
\medskip

\noindent {\bf Dependence on $x^a\in\R^{2,1}$.}  In Section~4 we have executed step (1) of the adiabatic approach algorithm and described the moduli space $\Mcal^{}_{S^1}$
of connections  $\Acal^{}_{S^1}$ on $S^1_R$. Now we return to Yang--Mills theory on $\R^{2,1}\times S^1_R$ as discussed in Section~2 and assume, according to step (2), 
that the gauge potential $\Acal =\Acal^{}_{\R^{2,1}}+\Acal^{}_{S^1}$ depends on  $x^a\in\R^{2,1}$ only via the coordinates $X^i$ on the moduli space $\Mcal^{}_{S^1}$, i.e.
\begin{equation}
X^i=X^i (x^a) \und \Acal_{\mu} =\Acal_{\mu}(X^i (x^a), x^3)\ .
\end{equation}
These moduli parameters $X=\{X^i\}$ define a map~\footnote{
not to be confused with the space~$X$}
\begin{equation}\label{5.1}
X:\quad \R^{2,1}\to G
\end{equation}
from $\R^{2,1}$ to the moduli space $\Mcal^{}_{S^1}\cong G$.
 
\medskip

\noindent {\bf Infinitesimal change of  $\Acal_3$.} For any fixed $x^a\in\R^{2,1}$, the part $\Acal^{}_{S^1}= \Acal^{}_{S^1}(X^i (x^a), x^3)$ of the gauge potential $ \Acal$
belongs to the space  $\Ncal^{}_{S^1}$, which is fibred (see  (\ref{4.5})) over the moduli space $G$ parametrized by  coordinates $X^i$. We introduce the tangent bundle
$T\Ncal^{}_{S^1}$ of $\Ncal^{}_{S^1}$ as a fibration
\begin{equation}\label{5.2}
h_{1\ast}:\quad T\Ncal^{}_{S^1}\to TG
\end{equation}
with fibres $T_{\Acal^{}_{S^1}}\O G\cong \O\gfrak$ at any point $\Acal^{}_{S^1}\in G$. For any given point $\Acal^{}_{S^1}\in G$ we have  $T_{\Acal^{}_{S^1}} G\cong \gfrak$ and therefore
\begin{equation}\label{5.3}
T_{\Acal^{}_{S^1}} \Ncal^{}_{S^1}\=h^*_1  T_{\Acal^{}_{S^1}} G\oplus T_{\Acal^{}_{S^1}}\O G\ \cong\ \gfrak\oplus\O\gfrak\ .
\end{equation}
Note that $x^a$ is an ``external'' parameter for  $\Acal^{}_{S^1}$ in  (\ref{5.2}) and  (\ref{5.3}), and the derivatives
\begin{equation}\label{5.4}
\pa_a \Acal^{}_{S^1} = \frac{\pa X^i}{\pa x^a}\, \pa_i\Acal^{}_{S^1} \quad\with\quad \pa_i=\frac{\pa}{\pa X^i}
\end{equation}
belong to the space $T_{\Acal^{}_{S^1}} \Ncal^{}_{S^1}$ for any $x^a\in\R^{2,1}$. According to  (\ref{5.3}), one can decompose the derivatives  (\ref{5.4})  into two parts,
\begin{equation}\label{5.5}
\pa_a \Acal_{3} \= (\pa_a X^i)\xi_{i3}+D_3(\eps_i{\pa_a X^i})\ ,
\end{equation}
where 
\begin{equation}\label{5.6}
\xi_{i3}\equiv\de_i \Acal_{3}
\end{equation}
belongs to $T_{\Acal^{}_{S^1}} G\cong \gfrak$ and $\eps_i$ belongs to $T_{\Acal^{}_{S^1}}\O G\cong \O\gfrak$, $i=1,\ldots, {\rm dim}G$. These $\eps_i$ are arbitrary 
$\gfrak$-valued gauge parameters, and 
\begin{equation}\label{5.7}
\eps_{a}:=\eps_{i}\pa_a X^i
\end{equation}
are their pull-back to $\R^{2,1}$.

It is natural to fix $\eps_{i}$ by requiring
\begin{equation}\label{5.8}
D_3\xi_{i3}=0\quad\Leftrightarrow\quad D^2_3\eps_i=D_3\pa_i\Acal_3
\end{equation}
so that $\xi_{i3}$ are orthogonal to infinitesimal gauge transformations of  $\Acal^{}_{S^1}$ generated by $\eps_{i}$. Note that these $\gfrak$-valued  gauge parameters $\eps_{i}$ 
define a connection $\eps_{i}\diff X^i$ on the moduli space  $\Mcal^{}_{S^1}$ (cf.~\cite{19, 21}), and $\eps_{a}$ from (\ref{5.7}) define a connection  $\eps_{a}\diff x^a$ on a 
$G$-bundle over $\R^{2,1}$ pulled back from  the connection $\eps_{i}\diff X^i$ on  $\Mcal^{}_{S^1}$.

\medskip

\noindent {\bf ``Electric'' part of effective action.} We discussed in detail the  $\Acal^{}_{S^1}$-part of the connection $\Acal =\Acal^{}_{\R^{2,1}}+\Acal^{}_{S^1}$ on $\R^{2,1}\times S^1_R$. 
On the other hand, the components $\Acal_a$ for $\Acal^{}_{\R^{2,1}}=\Acal_a\diff x^a$ are yet not fixed. It is natural to identify them with the sum of  $\eps_{a}$ from (\ref{5.7})~\cite{21} and 
arbitrary $\gfrak$-valued functions $\chi_a=\chi_a^i(X^j)\xi_{i3}$ which belong to the kernel of  $D_3$ due to   (\ref{5.8}),
\begin{equation}\label{5.9}
\Acal_a=\chi_a + \eps_a\qquad\Rightarrow\qquad \Fcal_{a3}=\pa_a\Acal_3 - D_3\Acal_a= (\pa_a X^i)\xi_{i3}\ \in T_{\Acal^{}_{S^1}}G\cong\gfrak\   .
\end{equation}
Both $\Acal_a\diff x^a$ and  $(\chi_a +\eps_{a})\diff x^a$ can be considered as gauge potentials on $\R^{2,1}$ with values in the loop algebra $L\gfrak = \gfrak\oplus\O\gfrak$.
Substituting  (\ref{5.9})  into  (\ref{2.6}), we obtain the term
\begin{equation}\label{5.10}
-\frac{1}{e^2 R}\int_{\R^{2,1}\times S^1} \!\!\!\!\diff^4x\ \eta^{ab}\ \tr\Fcal_{a3}\Fcal_{b3}\=
\frac{1}{e^2 R}\int_{\R^{2,1}} \!\diff^3x\ \eta^{ab}\  g_{ij}\ \pa_a X^i\pa_b X^j\ ,
\end{equation}
where 
\begin{equation}\label{5.11}
 g_{ij}= -\int_{S^1} \!\diff x^3\ \tr (\de_i\Acal_3\ \de_j\Acal_3)
\end{equation}
is a metric on the group $G$ in the holonomic basis. Thus, this part of the action  (\ref{2.6}) reduces to the action of a sigma model on $\R^{2,1}$ with target $\Mcal^{}_{S^1}\cong G$.

\medskip

\noindent {\bf ``Magnetic'' part of effective action.} Concerning the first term in the action (\ref{2.6}), the logic is as follows~\cite{3,4,5}. 
If the components $\Fcal_{ab}$ are nonsingular 
for $R\to 0$ then this term is negligible for small $R$ in comparison with the term  (\ref{5.10}), so it can be discarded. On the other hand, if we allow $R\Fcal_{ab}$ to remain
finite for $R\to 0$, then for the Abelian case $\Mcal^{}_{S^1}\cong T/W$ one can dualize to a ``magnetic'' photon~\cite{3,4,5}. 
In particular, the dual Abelian potential 
$\tilde{\Acal} = \tilde{\Acal}_{\mu}\diff x^{\mu}$ on $\R^{2,1}\times S^1$ is subject to
\begin{equation}\label{5.12}
\frac{1}{e}\ \Fcal_{ab}\= \frac{e}{R}\ \ve^{c}_{ab}\ \pa_c\tilde{\Acal}_3\ ,
\end{equation}
where the moduli space of $\tilde{\Acal}_3$ (the component of $\tilde{\Acal}$ along $S^1$) is parametrized by a dual torus $T^{\vee}$, 
which is a maximal torus in the Langlands dual group $G^{\vee}$.
Substituting (\ref{5.12}) into the action (\ref{2.6}) one generates the term
\begin{equation}\label{5.13}
\frac{e^2}{R}\int_{\R^{2,1}} \!\diff^3 x\ \eta^{ab}\ \tilde{g}_{\a\b}\ \pa_a  \tilde{X}^{\a}\pa_b  \tilde{X}^{\b}\ ,
\end{equation}
where $\tilde{X}^{\a}, \a =1,\ldots,r$, are coordinates on $T^{\vee}\subset {}G^{\vee}$
and $\tilde g_{\alpha\beta}$ is a metric on $T^{\vee}$. Of course, in this Abelian case in  (\ref{5.10}) one should keep only $X^{\a}\in T$. 

The action  (\ref{5.13}) can be generalized to the non-Abelian case. For this, let us admit a dual gauge potential  
$\tilde{\Acal} = \tilde{\Acal}_{a}\diff x^{a}+ \tilde{\Acal}_{3}\diff x^{3}$ taking values in the dual Lie algebra Lie$G^{\vee}$ and the corresponding dual gauge field
\begin{equation}\label{5.14}
\tilde\Fcal \= \sfrac12\tilde\Fcal_{\mu\nu}\diff x^\mu \wedge \diff x^\nu\quad\with\quad 
\tilde\Fcal_{\mu\nu} \=\partial_\mu\tilde\Acal_\nu - \partial_\nu\tilde\Acal_\mu + [\tilde\Acal_\mu , \tilde\Acal_\nu]\ ,
\end{equation}
where $\tilde\Fcal_{\mu\nu} =\tilde\Fcal_{\mu\nu} ^i\tilde I_i$ with generators~$\tilde I_i$ of $G^{\vee}$. Then to $\tilde{\Acal}_3$ and  $\tilde{\Fcal}_{a3}$
one can apply the same logic  as to ${\Acal}_3$ and  ${\Fcal}_{a3}$. We conclude that the moduli space of $\tilde{\Acal}_3$ living on $S^1$ 
is the dual Lie group $G^{\vee}$ and 
\begin{equation}\label{5.15}
\tilde\Fcal_{a3}\=(\partial_a\tilde{X}^i)\de_i\tilde{\Acal}_3\ ,
\end{equation}
where $\tilde{X}^i$ are local coordinates on $G^{\vee}$. The duality between $\Fcal_{\mu\nu}$ and $\tilde\Fcal_{\mu\nu}$ on $\R^{2,1}\times S^1$ is given by 
\begin{equation}\label{5.16}
\frac{1}{e}\ \Fcal_{\mu\nu}^i\= \frac{e}{2}\ \sqrt{|\det g^R|}\ \ve_{\mu\nu\lambda\sigma}\tilde\Fcal_R^{i\lambda\sigma}\ ,
\end{equation}
where $g_R^{\mu\nu}$ from (\ref{2.5}) is used for raising indices of $\tilde\Fcal_{\mu\nu}^i$. It follows from (\ref{5.16}) that
\begin{equation}\label{5.17}
\frac{1}{e}\ \Fcal_{ab}^i\= \frac{e}{R}\ \ve_{ab}^c\tilde\Fcal_{c3}^i\ .
\end{equation}
Using (\ref{5.15}) and (\ref{5.17}), we obtain 
\begin{equation}\label{5.18}
-\frac{1}{2e^2R}\int_{\R^{2,1}\times S^1} \!\!\!\!\diff^4x\,R^2\,\tr\Fcal_{ab}\Fcal^{ab}=
\frac{1}{4e^2 R}\int_{\R^{2,1}\times S^1} \!\!\!\!\diff^4x\,R^2\,\de_{kl}\Fcal_{ab}^k\Fcal^{lab}= 
\frac{e^2}{R}\,\int_{\R^{2,1}}\!\!\diff^3x\,\eta^{ab}\,\tilde g_{ij}\pa_a\tilde X^i\pa_b\tilde X^j\ ,
\end{equation}
where 
\begin{equation}\label{5.19}
\tilde g_{ij}\=\frac{1}{2}\int_{S^1}\diff x^3\, \de_{kl}\,(\de_i\tilde\Acal^k_3\ \de_j\tilde\Acal^l_3)   
\end{equation}
is a metric on the group $G^{\vee}$ in the holonomic basis. 
Thus, for small radius $R$ of the circle $S^1_R$ the Yang--Mills action on $\R^{2,1}\times S^1_R$ can be reduced to the effective action 
of a sigma model on $\R^{2,1}$ with target $G{\times}G^{\vee}$,
\begin{equation}\label{5.20}
S_{\sf eff}\=\frac{1}{R}\,\int_{\R^{2,1}}\!\!\diff^3x\,\Bigl(\frac{1}{e^2}\,\eta^{ab}\, g_{ij}\pa_a X^i\pa_b X^j + e^2\eta^{ab}\,\tilde g_{ij}\pa_a\tilde X^i\pa_b\tilde X^j\Bigr)\ .
\end{equation}
For $G=\;$SU($N$), this is the group SU$(N){\times}$SU$(N)/\Z_N$. For the Abelian case this action agrees with those considered in the literature.

\bigskip

\noindent {\bf Acknowledgements}

\noindent 
We thank Aleksey Cherman and Mohamed Anber for comments.
This work was partially supported by the Deutsche Forschungsgemeinschaft grant LE 838/13.
It is based upon work from COST Action MP1405 QSPACE, supported by COST (European Cooperation in Science and Technology).

%\newpage

\end{document}